\documentclass[12pt]{iopart}
\usepackage{graphicx}
\usepackage{dcolumn}
\usepackage{bm}
\usepackage{epsf}

\newcommand{\xiv}{\mbox{\boldmath$\xi$}}
\newcommand{\piv}{\mbox{\boldmath$\pi$}}
\newcommand{\sigv}{\mbox{\boldmath$\sigma$}}
\newcommand{\alphav}{\mbox{\boldmath$\alpha$}}

\begin{document}

\title{Kinematic mass of a composite in the many-particle Dirac model}
\author{P L Hagelstein$^1$, I U Chaudhary$^2$}

\address{$^1$ Research Laboratory of Electronics, 
Massachusetts Institute of Technology, 
Cambridge, MA 02139,USA}
\ead{plh@mit.edu}

\address{$^2$ 
Department of Computer Science and Engineering, 
University of Engineering and Technology, 
Lahore, Pakistan}
\ead{irfanc@mit.edu}

\begin{abstract}

We are interested in the energy-momentum relation for a moving composite in relativistic
quantum mechanics in many-particle Dirac models.
For a manifestly covariant model one can apply the Lorentz transform to go from the rest
frame to a moving frame to establish an energy-momentum relation of the form 
$\sqrt{(M^*c^2)^2+c^2|{\bf P}|^2}$
where $M^*$ is the kinematic mass.
However, the many-particle Dirac model is not manifestly covariant, and some other approach
is required.
We have found a simple approach that allows for a separation of relative and center of mass
contributions to the energy.  
We are able to define the associated kinematic energy and determine the energy-momentum
relation.
Our result can be expressed as a modified deBroglie relation of the form

$$
\hbar \omega ({\bf P})
~=~ 
\bigg \langle \Phi' \bigg |
\sum_j {m_j \over M} \beta_j   
\bigg | \Phi' \bigg \rangle
~
\sqrt{[M^*({\bf P}) c^2]^2 + c^2 |{\bf P}|^2}
$$

\noindent
where the kinematic mass $M^*$ will depend on the total momentum ${\bf P}$ for a general noncovariant
potential.  The prefactor that occurs we associate with a time dilation effect, the existence of
which has been discussed previously in the literature.

\end{abstract}

\submitto{\jpb}

\maketitle


\section{Introduction}
\label{sec:intro}

The many-particle Dirac model describes relativistic quantum mechanics of particles with 
spin 1/2.  
It is widely used for relativistic atomic, molecular, and nuclear structure calculations.
For relativistic kinematics, other relativistic quantum mechanical models derived from 
quantum field theory are much more widely used since they have been constructed
to be manifestly covariant.  
Although the many-particle Dirac model is useful for static problems,
it is not used very often for kinematic calculations involving many-particle composites
away from the rest frame.

A potential advantage of the many-electron Dirac model is that it is much simpler than quantum
field theory, the Bethe-Salpeter equation, and other modern relativistic models; 
we would like to be able to use it systematically to a wider class of problems than just static structure calculations.  
At issue is how the energy scales with the center of mass momentum, since the many-particle Dirac  
model is not manifestly covariant.
In this regard, consider the discussion given recently my Marsch \cite{Marsch2005}, in which the
two-body Dirac-Coulomb problem is solved in the rest frame; in the case of a hydrogen atom in
motion the energy is given as

\begin{equation}
E({\bf P}) ~=~ \sqrt{ E_0^2 + c^2 |{\bf P}|^2}
\label{eqn1}
\end{equation}

\noindent
where $E_0$ is the rest frame energy.  This is what we would expect in the case of a manifestly
covariant theory \cite{Salpeter1951}; however, the two-body Dirac-Coulomb model is not manifestly covariant, and 
it is not clear that we should expect this to come out as a result of a calculation done with
the two-body Dirac-Coulomb model.  A relativistic model need not exhibit a covariant energy
based on the kinematic center of mass energy, as demonstrated by Artru \cite{Artru1984}.
This point has been emphasized recently by Javenin \cite{Javenin2007}.

In a theory that is manifestly covariant one can take advantage of the Lorentz transform
to establish how the energy of a composite varies with the total momentum.  In a relativistic
quantum model that is not manifestly covariant, we would not expect a Lorentz boost of a wavefunction in
the rest frame to be consistent with a moving wavefunction in that model.  Consequently, 
in general we must use some other approach to determine the relation between the energy for a composite
at rest and in motion.

This problem is closely related to the separation of the center of mass and relative degrees of
freedom in relativistic quantum mechanics.  We would expect that in a consistent scheme in which
relative and center of mass degrees of freedom are separated, the center of mass problem
should be that of a free particle in the absence of an external potential.  Probably we would
be concerned if this were not the case in a particular relativistic quantum model. 

There is a substantial literature on two-body equations (and also for more complicated
systems) in relativistic quantum mechanics as well as quantum field theory, as discussed in 
\cite{Keister1991}, \cite{Malvetti1994}, \cite{Gross1999}, \cite{Pilkuhn2003}, and \cite{Javenin2007}. 

Our interest in the problem is motivated by issues rather different than those that have
been of concern to previous authors.  We have recently been interested in the impact of 
configuration mixing, or of an excited state superposition, of a composite on the center of 
mass dynamics.  The basic issue here is that if the different internal states have different energies,
then they have different masses, which should impact the ratio of momentum and velocity in the
nonrelativistic regime.  To study such problems we would like to use the many-particle Dirac model
as a foundation if possible (since it is simpler than field theory, and simpler than 
manifestly covariant models derived from field theory), since we require a relativistic model 
to describe the different masses associated with different excited states.  A model that
is approximately correct at low center of mass momentum is sufficient for us, which means we do not
require the model to be precise in the limit that the composite approaches the speed of light.
Finally, we do not require a manifestly covariant theory, but instead we would like an understandable
energy-momentum relation that has the correct (or at least understandable) scaling at low center of mass momentum.

We found a very simple approach that allows for a separation of the
center of mass and relative contributions to the eigenvalue in the many-particle Dirac model.
The resulting relation between eigenvalue and energy was unexpected (at least by us).
A part of the relation can clearly be associated with an energy-momentum relation of
the form of Equation (\ref{eqn1}), which allows for a consistent definition of the
kinematic mass.  There is in addition a prefactor, which may have physical significance
within the theory, in the context of a modified deBroglie relation.


\newpage
\section{Two-body Dirac model}

Although our approach is general, it is simplest to focus on the two-body 
version of the problem here.  The associated relative problem has been of interest 
since the early days of quantum mechanics \cite{Kemmer1937}, \cite{Fermi1949}, 
and the associated time-independent relative equation is sometimes called the 
Kemmer-Fermi-Yang equation \cite{Koide1982}, \cite{McNeil1992}.

\subsection{Hamiltonian}

The two-body Dirac Hamiltonian is

\begin{equation}
\hat{H}
~=~
\alphav_1 \cdot c\hat{\bf p}_1
+
\alphav_2 \cdot c\hat{\bf p}_2
+
\beta_1 m_1c^2
+
\beta_2 m_2c^2
+
V({\bf r}_2 - {\bf r}_1)
\end{equation}

\noindent
The $\alphav$ and $\beta$ matrices are 

\begin{equation}
\alphav 
~=~ 
\left (
\begin{array} {cc}
0 & \sigv \cr
\sigv & 0 \cr
\end{array}
\right )
\ \ \ \ \ \ \ \ \ \
\beta 
~=~ 
\left (
\begin{array} {cc}
I & 0 \cr
0 & -I \cr
\end{array}
\right )
\end{equation}

\noindent
where the $\sigv$ matrices are Pauli matrices.
Only relative forces are included in this model; the center of mass problem
here is that of a composite particle in free space.

\subsection{Classical center of mass and relative variables}

We define the classical center of mass and relative variables according to

\begin{equation}
M {\bf R} ~=~ m_1{\bf r}_1 + m_2 {\bf r}_2
\end{equation}

\begin{equation}
{\bf r} ~=~ {\bf r}_2 - {\bf r}_1
\end{equation}

\begin{equation}
{\bf P} ~=~ {\bf p}_1 + {\bf p}_2
\end{equation}

\begin{equation}
{{\bf p} \over \mu} ~=~ {{\bf p}_2 \over m_2} - {{\bf p}_1 \over m_1}
\end{equation}

\noindent
The total mass $M$ is

\begin{equation}
M ~=~ m_1 + m_2
\end{equation}

\noindent
and the relative mass $\mu$ satisfies

\begin{equation}
{1 \over \mu} ~=~ {1 \over m_1} + {1 \over m_2}
\end{equation}

\noindent
The classical position and momentum vectors are then

\begin{equation}
{\bf r}_1 ~=~ {\bf R} - {m_2 \over M}{\bf r} 
\ \ \ \ \ \ \ \ \ \ 
{\bf r}_2 ~=~ {\bf R} + {m_1 \over M} {\bf r} 
\end{equation}

\begin{equation}
{\bf p}_1 ~=~ {\mu \over m_2} {\bf P} - {\bf p} 
\ \ \ \ \ \ \ \ \ \ 
{\bf p}_2 ~=~ {\mu \over m_1} {\bf P}  + {\bf p} 
\end{equation}

\subsection{Hamiltonian in terms of center of mass and relative variables}

The Hamiltonian can be written in terms of the quantum momentum 
and relative position operators as

\begin{equation}
\hat{H}
~=~
\alphav_1 \cdot c \left ({\mu \over m_2} \hat{\bf P} - \hat{\bf p} \right )
+
\alphav_2 \cdot c \left ({\mu \over m_1} \hat{\bf P} + \hat{\bf p} \right )
+
\beta_1 m_1c^2
+
\beta_2 m_2c^2
+
V({\bf r})
\end{equation}

\noindent
This can be recast as

\begin{equation}
\hat{H}
~=~
\left ( {m_1 \over M} \alphav_1 + {m_2 \over M} \alphav_2 \right ) \cdot c \hat{\bf P} 
+
(\alphav_2-\alphav_1) \cdot c \hat{\bf p}
+
\beta_1 m_1c^2
+
\beta_2 m_2c^2
+
V({\bf r})
\end{equation}

\noindent
This is consistent with Barut and Stroebel \cite{Barut1986}.

\subsection{Time-dependent problem and eigenvalue equation}

The time-dependent two-body Dirac equation is

\begin{equation}
i \hbar {\partial \over \partial t} \Psi({\bf r}_1,{\bf r}_2,t)
~=~=
\hat{H}
\Psi({\bf r}_1,{\bf r}_2,t)
\end{equation}

\noindent
We assume a stationary state solution of the form

\begin{equation}
\Psi({\bf r}_1,{\bf r}_2,t)
~=~
e^{-i \omega t} \Phi({\bf r}_1,{\bf r}_2)
\end{equation}

\noindent
where $\Phi({\bf r}_1,{\bf r}_2)$ satisfies the eigenvalue equation 

\begin{equation}
\hbar \omega \Phi({\bf r}_1,{\bf r}_2) ~=~ \hat{H} \Phi({\bf r}_1,{\bf r}_2)
\end{equation}


\newpage
\section{Center of mass and relative contributions to the eigenvalue}

In the nonrelativistic problem, we are able to split the Hamiltonian cleanly into
center of mass and relative pieces.  Although there is no equivalent separation in 
relativistic quantum mechanics, it appears to be possible to develop a simple separation
of the associated Hamiltonians (as we discuss in this section).  

We note that Barut and coworkers have discussed the separation of relative and center of mass
degrees of freedom for this problem \cite{Barut1986}, \cite{Barut1985a}, \cite{Barut1985}.
Of the relevant papers in the literature, our approach is most closely related to 
the ideas in these papers; however, our conclusions are quite different.

\subsection{Eigenvalue equation and expectation values}

Consider now the time-independent two-body Dirac equation

\begin{equation}
\hbar \omega \Phi
~=~
\bigg [
\left ( {m_1 \over M} \alphav_1 + {m_2 \over M} \alphav_2 \right ) \cdot c \hat{\bf P} 
+
(\alphav_2-\alphav_1) \cdot c \hat{\bf p}
+
\beta_1 m_1c^2
+
\beta_2 m_2c^2
+
V({\bf r})
\bigg ] \Phi
\end{equation}

\noindent
Suppose now that $\Phi$ is an exact solution.  If so, then the eigenvalue may
be expressed in terms of expectation values according to

$$
\hbar \omega
~=~
\bigg \langle \Phi \bigg |
\left ( {m_1 \over M} \alphav_1 + {m_2 \over M} \alphav_2 \right ) \cdot c \hat{\bf P} 
\bigg | \Phi \bigg \rangle
\ \ \ \ \ \ \ \ \ \ \ \ \ \ \ \ \ \ \ \ 
\ \ \ \ \ \ \ \ \ \ \ \ \ \ \ \ \ \ \ \ 
$$
\begin{equation}
\ \ \ \ \ \ \ \ \ \ 
+
\bigg \langle \Phi \bigg |
(\alphav_2-\alphav_1) \cdot c \hat{\bf p}
+
\beta_1 m_1c^2
+
\beta_2 m_2c^2
+
V({\bf r})
\bigg |\Phi \bigg \rangle
\end{equation}

\subsection{Kinematic mass}

\noindent
Suppose that we now add and subtract kinematic mass terms to the expectation values; this allows us to write

$$
\hbar \omega
~=~
\bigg \langle \Phi \bigg |
\left ( {m_1 \over M} \alphav_1 + {m_2 \over M} \alphav_2 \right ) \cdot c \hat{\bf P} 
+
\left ( {\mu \over m_2} \beta_1 + {m_2 \over M} \beta_2 \right )M^* c^2 
\bigg | \Phi \bigg \rangle
\ \ \ \ \ \ \ \ \ \ \ \ \ \ \ \ \ \ \ \ 
$$
\begin{equation}
+
\bigg \langle \Phi \bigg |
(\alphav_2-\alphav_1) \cdot c \hat{\bf p}
+
\beta_1 m_1c^2
+
\beta_2 m_2c^2
+
V({\bf r})
-
\left ( {\mu \over m_2} \beta_1 + {m_2 \over M} \beta_2 \right )M^* c^2 
\bigg |\Phi \bigg \rangle
\end{equation}

\noindent
If we assume that the contribution of the kinematic mass to the energy appears
only through the first term, then the second term must vanish

\newpage

\begin{equation}
\bigg \langle \Phi \bigg |
(\alphav_2-\alphav_1) \cdot c \hat{\bf p}
+
\beta_1 m_1c^2
+
\beta_2 m_2c^2
+
V({\bf r})
-
\left ( {m_1 \over M} \beta_1 + {m_2 \over M} \beta_2 \right )M^* c^2 
\bigg |\Phi \bigg \rangle
~=~
0
\end{equation}

\noindent
This allows us to define the kinematic mass from the raio

\begin{equation}
M^* c^2
~=~
{
\displaystyle{
\bigg \langle \Phi \bigg |
(\alphav_2-\alphav_1) \cdot c \hat{\bf p}
+
\beta_1 m_1c^2
+
\beta_2 m_2c^2
+
V({\bf r})
\bigg |\Phi \bigg \rangle
}
\over
\displaystyle{
\bigg \langle \Phi \bigg |
 {m_1 \over M} \beta_1 + {m_2 \over M} \beta_2 
\bigg |\Phi \bigg \rangle
}
}
\end{equation}

\noindent
Note that it is possible that the kinematic mass defined in this
way will have different values for different choices of the center of
mass momentum, since the two-body Dirac model with arbitrary $V({\bf r})$ is not manifestly covariant.

\subsection{Eigenvalue and energy}

With the kinematic mass defined in this way, the eigenvalue then simplifies to

\begin{equation}
\hbar \omega
~=~
\bigg \langle \Phi \bigg |
\left ( {m_1 \over M} \alphav_1 + {m_2 \over M} \alphav_2 \right ) \cdot c \hat{\bf P} 
+
\left ( {m_1 \over M} \beta_1 + {m_2 \over M} \beta_2 \right )M^* c^2 
\bigg | \Phi \bigg \rangle
\end{equation}

\noindent
We can rotate to obtain

\begin{equation}
\hbar \omega
~=~
\bigg \langle \Phi' \bigg |
\left ( {m_1 \over M} \beta_1 + {m_2 \over M} \beta_2 \right ) \sqrt{ (M^* c^2)^2 + c^2 |\hat{\bf P}|^2} 
\bigg | \Phi' \bigg \rangle
\end{equation}

\noindent
If we assume that $\Phi'$ has a plane-wave dependence on the center of mass coordinate

\begin{equation}
\Phi' ~\sim~ e^{i {\bf P}\cdot {\bf R}/\hbar}
\end{equation}

\noindent
then we may simplify to

\begin{equation}
\hbar \omega
~=~
\bigg \langle \Phi' \bigg |
 {m_1 \over M} \beta_1 + {m_2 \over M} \beta_2  
\bigg | \Phi' \bigg \rangle
\sqrt{ (M^* c^2)^2 + c^2 |{\bf P}|^2}
\end{equation}

We recall that the relativistic energy for a moving composite is

\begin{equation}
E({\bf P})
~=~
\sqrt{ (M^* c^2)^2 + c^2 |{\bf P}|^2}
\end{equation}

\noindent
with a fixed kinematic mass.
We see that the eigenvalue $\hbar \omega$ is proportional to the energy-momentum
relation, but a prefactor is evident.  We may write

\begin{equation}
\hbar \omega ({\bf P})
~=~
\bigg \langle \Phi'({\bf P}) \bigg |
 {m_1 \over M} \beta_1 + {m_2 \over M} \beta_2  
\bigg | \Phi'({\bf P}) \bigg \rangle
\sqrt{ [M^*({\bf P}) c^2]^2 + c^2 |{\bf P}|^2}
\end{equation}

\subsection{Rest frame kinematic mass}

Bound state computations are usually performed in the rest frame, which suggests
that the kinematic mass in the rest frame (where probably the potential model will
be most relevant) should be estimated first by solving

\begin{equation}
\hbar \omega
\Phi 
~=~
\bigg [
(\alphav_2-\alphav_1) \cdot c \hat{\bf p}
+
\beta_1 m_1c^2
+
\beta_2 m_2c^2
+
V({\bf r})
\bigg ]
\Phi 
\end{equation}

\noindent
and then evaluating

\begin{equation}
M^* c^2
~=~
{
\displaystyle{
\bigg \langle \Phi \bigg |
(\alphav_2-\alphav_1) \cdot c \hat{\bf p}
+
\beta_1 m_1c^2
+
\beta_2 m_2c^2
+
V({\bf r})
\bigg |\Phi \bigg \rangle
}
\over
\displaystyle{
\bigg \langle \Phi \bigg |
 {m_1 \over M} \beta_1 + {m_2 \over M} \beta_2 
\bigg |\Phi \bigg \rangle
}
}
\bigg |_{{\bf P}=0}
\end{equation}

\subsection{Extension to many particles}

This approach can be extended to the case of many Dirac particles, as discussed 
in the Appendix.  The eigenvalue in this case is of the form

\begin{equation}
\hbar \omega
~=~
\bigg \langle \Phi' \bigg |
\sum_j {m_j \over M} \beta_j   
\bigg | \Phi' \bigg \rangle
\sqrt{(M^* c^2)^2 + c^2 |{\bf P}|^2}
\end{equation}

\noindent
where the kinematic mass is

\begin{equation}
M^*c^2
~=~
{
\displaystyle{
\bigg \langle \Phi \bigg |
\sum_j \alphav_j \cdot c \hat{\piv}_j 
+
\sum_j \beta_j m_jc^2
+
\sum_{j<k} V_{jk}(\xiv_k-\xiv_j)
\bigg | \Phi \bigg \rangle
}
\over
\displaystyle{
\bigg \langle \Phi \bigg |
\sum_j {m_j \over M} \beta_j 
\bigg | \Phi \bigg \rangle
}
}
\end{equation}


\newpage
\section{Associated deBroglie relation}

From the discussion above, we have found that the many-particle Dirac equation
leads us to a deBroglie relation of the form

$$
\hbar \omega
~=~
\bigg \langle \Phi' \bigg |
\sum_j {m_j \over M} \beta_j   
\bigg | \Phi' \bigg \rangle
\sqrt{(M^* c^2)^2 + c^2 |{\bf P}|^2}
$$

\noindent
In light of the energy-momentum relation from the Lorentz transform

$$E({\bf P}) ~=~ \sqrt{(M^* c^2)^2 + c^2 |{\bf P}|^2}$$

\noindent
we might view this result as consistent with a modified deBroglie relation 

\begin{equation}
\hbar \omega ~=~ 
\bigg \langle \Phi' \bigg |
\sum_j {m_j \over M} \beta_j   
\bigg | \Phi' \bigg \rangle
E
\end{equation}

It might be asked why such a prefactor should occur?  After some thought, it seems
that we might associate it with a time-dilation effect associated with the potential
in the case of a bound composite.  There is precedent for such an effect in the
literature (see \cite{Holten1991}, \cite{Holten1992}).


\newpage
\section{Discussion and conclusion}

This study was motivated by our interest in problems where the internal energy
of a composite impacts the kinematics and dynamics at low center of mass momentum.
The many-particle Dirac model is perhaps the simplest relativistic quantum
mechanical model that might be relevant, but it has not been used much in the literature
for problems outside of the rest frame.  Since the many-particle Dirac model in
general is not manifestly covariant, we are not able to use a Lorentz transformation to understand
how the energy scales with momentum.

We found a simple approach which allows us to separate the center of mass and relative
contributions to the eigenvalue.  The resulting eigenvalue relation in general can be written  
in the form 

\begin{equation}
\hbar \omega ({\bf P})
~=~ 
\bigg \langle \Phi' \bigg |
\sum_j {m_j \over M} \beta_j   
\bigg | \Phi' \bigg \rangle
~
\sqrt{[M^*({\bf P}) c^2]^2 + c^2 |{\bf P}|^2}
\end{equation}

\noindent
where we have identified $M^*$ as a kinematic mass (which we would expect to be independent of ${\bf P}$ in a covariant
many-particle Dirac model). 

If we view the eigenvalue relation as a modified deBroglie relation

$$
\hbar \omega ~=~ 
\bigg \langle \Phi' \bigg |
\sum_j {m_j \over M} \beta_j   
\bigg | \Phi' \bigg \rangle
E
$$

\noindent
then the question to be faced concerns the interpretation of the prefactor.  It seems that
we might best associate this with a time dilation effect, consistent with recent
literature predicting such an effect.


\newpage
\appendix

\section{Kinematic mass in the many-particle Dirac model}

The arguments given in the main text for the two-body problem can be extended to
more Dirac particles directly, as we discuss below.

\subsection{Classical center of mass and relative coordinates}

We begin by defining the classical center of mass coordinate through

\begin{equation}
M {\bf R} ~=~ \sum_j m_j {\bf r}_j
\end{equation}

\noindent
where

\begin{equation}
M  ~=~ \sum_j m_j 
\end{equation}

\noindent
Relative coordinates can be defined according to

\begin{equation}
\xiv_j ~=~ {\bf r}_j - {\bf R}
\end{equation}

\noindent
If there are $N$ particles, then we may define $N$ relative coordinates,
but one of them is redundant because

\begin{equation}
\sum_j \xiv_j ~=~ 0
\end{equation}

\subsection{Classical momentum coordinates}

The total classical momentum is the sum of the individual particle momenta

\begin{equation}
{\bf P} ~=~ \sum_j {\bf p}_j
\end{equation}

\noindent
The relative momenta are defined according to

\begin{equation}
\piv_j ~=~ {\bf p}_j - {m_j \over M} {\bf P}
\end{equation}

\noindent
Once again there are more relative momenta than required, since

\begin{equation}
\sum_j \piv_j ~=~ \sum_j {\bf p}_n - {{\bf P} \over M} \sum_j m_j ~=~ 0
\end{equation}

\subsection{Many-particle Dirac Hamiltonian}

The many-particle Dirac Hamiltonian is

\begin{equation}
\hat{H}
~=~
\sum_j \alphav_j \cdot c \hat{\bf p}
+
\sum_j \beta_j m_jc^2
+
\sum_{j<k} V_{jk}({\bf r}_k-{\bf r}_j)
\end{equation}

\noindent
where we presume that the potentials $V_{jk}$ are functions of relative coordinates.  We can
recast this in terms of center of mass and relative coordinates to obtain

\begin{equation}
\hat{H}
~=~
\sum_j \alphav_j \cdot c \bigg ( \hat{\piv}_j + {m_j \over M} \hat{\bf P} \bigg )
+
\sum_j \beta_j m_jc^2
+
\sum_{j<k} V_{jk}(\xiv_k-\xiv_j)
\end{equation}

\subsection{Solution and expectation values}

We assume that $\Phi$ satisfies the time-independent Dirac equation

\begin{equation}
\hbar \omega \Phi
~=~
\hat{H} \Phi
\end{equation}

\noindent
The eigenvalue then can be expressed in terms of the wavefunction as

$$
\hbar \omega
~=~
\bigg \langle \Phi \bigg |
\left ( \sum_j {m_j \over M} \alphav_j \right ) \cdot c   \hat{\bf P} 
\bigg | \Phi \bigg \rangle
\ \ \ \ \ \ \ \ \ \ \ \ \ \ \ \ \ \ \ \ \ \ \ \ \ \ \ \ \ \ 
\ \ \ \ \ \ \ \ \ \ \ \ \ \ \ \ \ \ \ \ \ \ \ \ \ \ \ \ \ \ 
$$
\begin{equation}
+
\bigg \langle \Phi \bigg |
\sum_j \alphav_j \cdot c \hat{\piv}_j 
+
\sum_j \beta_j m_jc^2
+
\sum_{j<k} V_{jk}(\xiv_k-\xiv_j)
\bigg | \Phi \bigg \rangle
\end{equation}

\subsection{Kinematic mass}

We can add kinematic mass terms and write

$$
\hbar \omega
~=~
\bigg \langle \Phi \bigg |
\left ( \sum_j {m_j \over M} \alphav_j \right ) \cdot c   \hat{\bf P} 
+
\left ( \sum_j {m_j \over M} \beta_j \right ) M^* c^2 
\bigg | \Phi \bigg \rangle
\ \ \ \ \ \ \ \ \ \ \ \ \ \ \ \ \ \ \ \ \ \ \ \ \ \ \ \ \ \ 
\ \ \ \ \ \ \ \ \ \ \ \ \ \ \ \ \ \ \ \ \ \ \ \ \ \ \ \ \ \ 
$$
\begin{equation}
+
\bigg \langle \Phi \bigg |
\sum_j \alphav_j \cdot c \hat{\piv}_j 
+
\sum_j \beta_j m_jc^2
+
\sum_{j<k} V_{jk}(\xiv_k-\xiv_j)
-
\left ( \sum_j {m_j \over M} \beta_j \right ) M^* c^2 
\bigg | \Phi \bigg \rangle
\end{equation}

\noindent
Once again we require that the eigenvalue have no contribution from the relative problem, which
occurs if we compute the kinematic mass according to 

\begin{equation}
M^*c^2
~=~
{
\displaystyle{
\bigg \langle \Phi \bigg |
\sum_j \alphav_j \cdot c \hat{\piv}_j 
+
\sum_j \beta_j m_jc^2
+
\sum_{j<k} V_{jk}(\xiv_k-\xiv_j)
\bigg | \Phi \bigg \rangle
}
\over
\displaystyle{
\bigg \langle \Phi \bigg |
\sum_j {m_j \over M} \beta_j 
\bigg | \Phi \bigg \rangle
}
}
\end{equation}

\subsection{Eigenvalue and energy}

The eigenvalue for the many particle Dirac problem in this case is

\begin{equation}
\hbar \omega
~=~
\bigg \langle \Phi \bigg |
\left ( \sum_j {m_j \over M} \alphav_j \right ) \cdot c   \hat{\bf P} 
+
\left ( \sum_j {m_j \over M} \beta_j \right ) M^* c^2 
\bigg | \Phi \bigg \rangle
\end{equation}

\noindent
We can rotate to obtain

\begin{equation}
\hbar \omega
~=~
\bigg \langle \Phi' \bigg |
\sum_j {m_j \over M} \beta_j   
\bigg | \Phi' \bigg \rangle
\sqrt{(M^* c^2)^2 + c^2 |{\bf P}|^2}
\end{equation}

\noindent
assuming that

\begin{equation}
\Phi' 
~\sim~
e^{i {\bf P} \cdot {\bf R} / \hbar}
\end{equation}


\end{document}